\documentclass[preprint,aps,showpacs,nofootinbib,preprintnumbers,amsmath,amssymb]{revtex4-1}
\usepackage{}
\usepackage{epsfig}
\usepackage{subfigure}
\usepackage{dcolumn}
\usepackage{bm}
\usepackage[usenames ,dvipsnames]{xcolor}
\usepackage{slashed}
\usepackage{graphicx,color}

\begin{document}

\title{Violation of PCAC and two-body baryonic $B$ and $D_s$ decays}
\author{Y.K. Hsiao$^{a,b}$ and C.Q. Geng$^{a,b,c}$}
\affiliation{
$^a$Physics Division, National Center for Theoretical Sciences, Hsinchu 300, Taiwan\\
$^b$Department of Physics, National Tsing Hua University, Hsinchu 300, Taiwan\\
$^c$Chongqing University
of Posts \& Telecommunications, Chongqing, 400065, China}

\begin{abstract}
We study the two-body baryonic $B$ and $D_s$ decays based on the annihilation mechanism
without the partial conservation of axial-vector current (PCAC)  at the GeV scale. We demonstrate
that the contributions of $B^-\to \Lambda\bar p$, $B^-\to \Sigma^0 \bar p$ and
$\bar B^0_s\to \Lambda\bar \Lambda$ are mainly
from the scalar and pseudoscalar currents with their
branching ratios  predicted to be around
$(3.5,\,5.3,\,5.3)\times 10^{-8}$, respectively,
exactly the sizes of ${\cal B}(B\to {\bf B\bar B'})$ established by the data.
We also apply the annihilation mechanism  to all of the charmless two-body baryonic
$B$ and $D_s$ decays. In particular, we can explain ${\cal B}(\bar B^0_{(s)}\to p\bar p)$
of order $10^{-8}$ and ${\cal B}(D_s^+\to p\bar n)$ of order $10^{-3}$, which are from
the axial-vector currents. In addition, the branching ratios of
$\bar B^0\to \Lambda\bar \Lambda$, $B^-\to n\bar p$, and
$B^-\to \Sigma^- \bar\Sigma^0$ are predicted to be
$(0.3,\,3.2,\,9.6)\times 10^{-8}$,
which can be measured by LHCb and viewed as
tests for the violation of PCAC at the GeV scale.
\end{abstract}


\maketitle

\section{Introduction}
For the abundantly observed three-body baryonic $B$ decays ($B\to {\bf B\bar B'}M$),
the theoretical approach for the systematic study has been
established~\cite{HouSoni,Cheng:2001tr,ChuaHou,ChuaHouTsai,AngdisppK,NF_GengHsiao}.
It leads to the theoretical predictions, among which at least five decay modes~\cite{GengHsiao5,GengHsiaoHY}
are observed to agree with the data~\cite{pdg}.
On the other hand,  the two-body baryonic $B$ decays ($B\to {\bf B\bar B'}$) are poorly understood
due to the smaller branching ratios, causing a much later observation than $B\to {\bf B\bar B'}M$.
Recently, the LHCb collaboration has presented the first observations of the charmless
$B\to {\bf B\bar B'}$ decays~\cite{Aaij:2013fta},  given by
\begin{eqnarray}\label{data1}
{\cal B}(\bar B^0\to p\bar p)&=&(1.47^{+0.62+0.35}_{-0.51-0.14})\times 10^{-8}\,,\nonumber\\
{\cal B}(\bar B^0_s\to p\bar p)&=&(2.84^{+2.03+0.85}_{-1.68-0.18})\times 10^{-8}\,,
\end{eqnarray}
with the statistical significances to be $3.3\sigma$ and $1.9\sigma$, respectively.

Based on the factorization, 
when the $B$ meson annihilates with the momentum transfer $q$,
the amplitudes ${\cal A}(\bar B^0_{(s)}\to p\bar p)$ can be decomposed as
$q^\mu\langle p\bar p|A_\mu|0\rangle$, where the matrix element is for
the proton pair production and $A_\mu$ is the axial-vector current.
 From the hypothesis of the partial conservation of the axial-vector
 current (PCAC)~\cite{PCAC} at the GeV scale,
 $q^\mu A_\mu$ is proportional to $m_\pi^2$,
which leads to ${\cal A}(\bar B^0_{(s)}\to p\bar p)\simeq 0$.
This is the reason why the non-factorizable
effects were believed to dominate the branching ratios
in Eq.~(\ref{data1})~\cite{diagramic1}\footnote{For the review of the various models, please consult Ref.~\cite{diagramic1},
and the references therein.}.
However, since the predictions from these models differ from each other, and commonly exceed
the data, a reliable theoretical approach has not been established yet.

In this work, we would propose a new method without the use of PCAC. In fact, the smallness
of the previous estimations is not caused by the annihilation mechanism~\cite{Pham:1980dc},
but the assumption of PCAC. Moreover, this assumption has never been tested  at the GeV scale.
For example, ${\cal B}(B^-\to\Lambda\bar p)$ and ${\cal B}(\bar B^0_s\to\Lambda\bar \Lambda)$
are found to have the amplitudes decomposed as
$({m_B^2}/{m_b}) \langle p\bar p|S+P|0\rangle$ with $S(P)$ the (pseudo)scalar current,
which has no connection to PCAC. Since they can be estimated to be of order $10^{-8}$,
exactly the order of the magnitude of ${\cal B}(B\to{\bf B\bar B'})$ measured by the experiments,
the annihilation mechanism can be justified.
If the axial-vector current is asymptotically conserved,
the result of ${\cal B}(D_s^+\to p\bar n)=(0.4^{+1.1}_{-0.3})\times 10^{-6}$ in Ref.~\cite{Chen:2008pf}
would yield ${\cal B}(D_s^+\to p\bar n)/{\cal B}(D_s^+\to \tau\bar \nu_\tau)\simeq 10^{-5}$,
which was indeed suggested as the test of PCAC at the GeV scale~\cite{Pham:1980dc}.
Nonetheless, with
${\cal B}(D_s^+\to p\bar n)$$=(1.30\pm 0.36^{+0.12}_{-0.16})\times 10^{-3}$ measured
by the CLEO Collaboration~\cite{Athar:2008ug},
one obtains that ${\cal B}(D_s^+\to p\bar n)/{\cal B}(D_s^+\to \tau\bar \nu_\tau)\simeq 0.02$,
which is too large and can be viewed as
a counter case of PCAC~\cite{Bediaga:1991eu}.

In this paper,
we apply the annihilation mechanism  to the two-body baryonic $B$ decays,
provided that the axial-vector current is not asymptotically conserved.
By modifying the timelike baryonic form factors via the axial-vector current without respect to PCAC,
we can explain ${\cal B}(\bar B^0_{(s)}\to p\bar p)$ as well as ${\cal B}(D_s^+\to p\bar n)$.
We shall also predict
${\cal B}(B^-\to\Lambda(\Sigma^0)\bar p)$ and ${\cal B}(\bar B^0_s\to\Lambda\bar \Lambda)$
in terms of the timelike baryonic form factors via the scalar and pseudoscalar currents.

The paper is organized as follows. In Sec.~2, we present the formalism of  the two-body baryonic $B$ and $D_s$ decays.
In Sec.~3, we  proceed  our numerical analysis.
  Sec.~4 contains our discussions and conclusions.

\section{Formalism}
\begin{figure}[t!]
\centering
\includegraphics[width=2.0in]{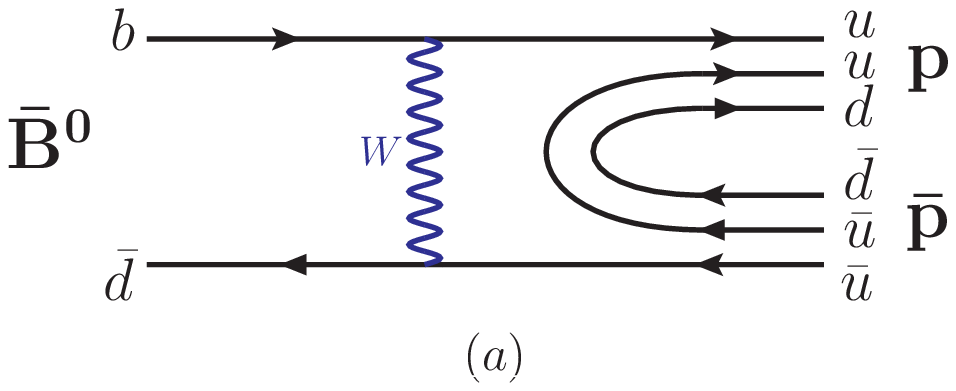}
\includegraphics[width=2.0in]{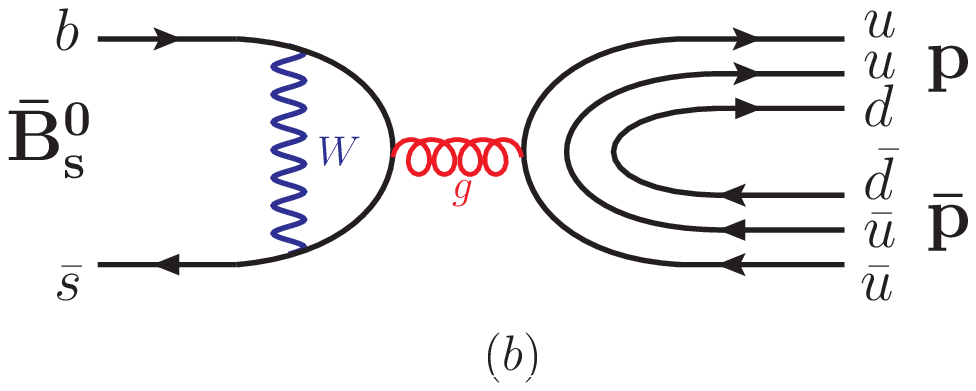}
\includegraphics[width=2.0in]{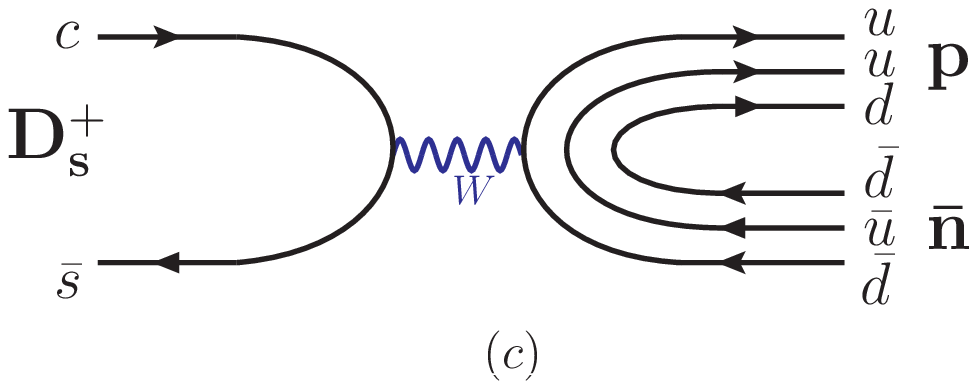}
\caption{The two-body baryonic decays of
(a)$\bar B^0\to p\bar p$,
(b)$\bar B^0_s\to p\bar p$,
and (c)$D^+_s\to p\bar n$.}\label{figBtoBB}
\end{figure}
The non-leptonic $B$ and $D$ decays in the factorization hypothesis
are in analogy with the semileptonic cases like
${\cal A}(B\to \pi e\bar \nu_e)\propto
\langle \pi|u\gamma^\mu(1-\gamma_5)b|B\rangle \bar e\gamma_\mu(1-\gamma_5)\nu_e$
to have the amplitudes with an additional matrix element in the form of
$\langle X_2|J^{2(\mu)}|0\rangle \langle X_1|J^1_{(\mu)}|B\rangle$,
where $J^{1,2}_{(\mu)}$ are the quark currents, and
$X_{1,2}$ can be multi-hadron states~\cite{fac1,fac2}.
Although the derivation may not be analytically satisfactory,
the factorization approximation can still be justified
by theoretically reproducing the data and predicting not-yet-observed decay modes
to be approved by the later measurements
in the two-body and three-body mesonic $B$ decays
as well as the three-body baryonic $B$ decays~\cite{GengHsiaoHY,ali,Hamiltonian,3b_HY}.

Like the measured $\bar B^0_{(s)}\to p\bar p$ and $D_s^+\to p\bar n$
with the decaying processes depicted in Fig.~\ref{figBtoBB},
in the two-body baryonic $B$ and $D_s$ decays,
the factorizable amplitudes are known to depend on
the annihilation mechanism~\cite{Pham:1980dc,Bediaga:1991eu},
where $B$ and $D_s$ annihilate, followed by the baryon pair production.
Thus, the amplitudes can have two types, ${\cal A}_1$ and ${\cal A}_2$,
which consist of (axial)vectors and (pseudo)scalar quark currents, respectively.
For example, the amplitudes of $\bar B^0\to (p\bar p$, $\Lambda\bar \Lambda)$,
$B^-\to (n\bar p$, $\Sigma^-\bar \Sigma^0)$,
and $D_s^+\to p\bar n$ are of the first type,
given by~\cite{Pham:1980dc,Bediaga:1991eu,Chen:2008pf}
\begin{eqnarray}\label{amp1}
&&{\cal A}_1(\bar B^0\to {\bf B_1\bar B_1'})=\frac{G_F}{\sqrt 2}
V_{ub}V_{ud}^*\,a_2\langle{\bf B_1\bar B_1'}|(\bar u u)_{V-A}|0\rangle
\langle 0|(\bar d b)_{V-A}|\bar B^0\rangle\,, \nonumber\\
&&{\cal A}_1(B^-\to {\bf B_2\bar B_2'})=\frac{G_F}{\sqrt 2}
V_{ub}V_{ud}^*\,a_1\langle {\bf B_2\bar B_2'}|(\bar d u)_{V-A}|0\rangle
\langle 0|(\bar u b)_{V-A}|B^-\rangle\,, \nonumber\\
&&{\cal A}_1(D_s^+\to p\bar n)=\frac{G_F}{\sqrt 2}
V_{cs}V_{ud}^*\,a_1\langle p\bar n|(\bar u d)_{V-A}|0\rangle
\langle 0|(\bar s c)_{V-A}|D^+_s\rangle\,,
\end{eqnarray}
where ${\bf B_1\bar B_1'}=p\bar p$ or $\Lambda\bar \Lambda$,
${\bf B_2\bar B_2'}=n\bar p$ or $\Sigma^-\bar \Sigma^0$,
$(\bar q_1 q_2)_{V-A}$ denotes $\bar q_1\gamma_\mu(1-\gamma_5) q_2$,
 $G_F$ is the Fermi constant, $a_{i}$ are the effective Wilson coefficients,
and  $V_{q_1q_2}$ are the Cabibbo-Kobayashi-Maskawa (CKM) matrix elements.
The amplitudes of $\bar B^0_s\to (p\bar p$, $\Lambda\bar \Lambda)$ and
$B^-\to (\Lambda\bar p$, $\Sigma^0\bar p)$ are more complicated, written as
\begin{eqnarray}\label{amp1a}
&&{\cal A}(\bar B^0_s\to {\bf B_1\bar B_1'})=
{\cal A}_1(\bar B^0_s\to {\bf B_1\bar B_1'})+{\cal A}_2(\bar B^0_s\to {\bf B_1\bar B_1'})\,,
\nonumber\\
&&{\cal A}(B^-\to {\bf B_2\bar B_2'})=
{\cal A}_1(B^-\to {\bf B_2\bar B_2'})+{\cal A}_2(B^-\to {\bf B_2\bar B_2'})\,,
\end{eqnarray}
where
\begin{eqnarray}\label{amp1b}
{\cal A}_1(\bar B^0_s\to {\bf B_1\bar B_1'}) &=&\frac{G_F}{\sqrt 2}\bigg\{
V_{ub}V_{us}^*\,a_2\langle {\bf B_1\bar B_1'}|(\bar u u)_{V-A}|0\rangle
\nonumber\\&&
-V_{tb}V_{ts}^*\bigg[a_3\langle {\bf B_1\bar B_1'}|(\bar u u+\bar d d+\bar s s)_{V-A}|0\rangle
\nonumber\\
&&+a_4\langle{\bf B_1\bar B_1'}|(\bar s s)_{V-A}|0\rangle+
a_5\langle {\bf B_1\bar B_1'}|(\bar u u+\bar d d+\bar s s)_{V+A}|0\rangle
\nonumber\\&&
+\frac{a_9}{2}\langle {\bf B_1\bar B_1'}|(2\bar u u-\bar d d-\bar s s)_{V-A}|0\rangle\bigg]\bigg\}
\langle 0|(\bar s b)_{V-A}|\bar B^0_s\rangle\,,
\nonumber\\
{\cal A}_1(B^-\to {\bf B_2\bar B_2'})&=&\frac{G_F}{\sqrt 2}
(V_{ub}V_{us}^*\,a_1-V_{tb}V_{ts}^*\,a_4)
\langle {\bf B_2\bar B_2'}|(\bar s u)_{V-A}|0\rangle
\langle 0|(\bar u b)_{V-A}|B^-\rangle\,,~~
\end{eqnarray}
and
\begin{eqnarray}\label{amp2}
{\cal A}_2(\bar B^0_s\to {\bf B_1\bar B_1'})&=&\frac{G_F}{\sqrt 2}V_{tb}V_{ts}^*\,2a_6
\langle{\bf B_1\bar B_1'}|(\bar s s)_{S+P}|0\rangle
\langle 0|(\bar s b)_{S-P}|\bar B^0_s\rangle\,,
\nonumber\\
{\cal A}_2(B^-\to {\bf B_2\bar B_2'})&=&\frac{G_F}{\sqrt 2}V_{tb}V_{ts}^*\,2a_6
\langle{\bf B_2\bar B_2'}|(\bar s u)_{S+P}|0\rangle
\langle 0|(\bar u b)_{S-P}|B^-\rangle\,,
\end{eqnarray}
with ${\bf B_1\bar B_1'}=p\bar p$ or $\Lambda\bar \Lambda$,
${\bf B_2\bar B_2'}=\Lambda\bar p$ or $\Sigma^0\bar p$, and $(\bar q_1 q_2)_{S\pm P}$
representing $\bar q_1(1\pm\gamma_5) q_2$. For the coefficients $a_i$ in Eqs.~(\ref{amp1})-(\ref{amp2}),
we use the same inputs as those in $B\to{\bf B\bar B'}M$~\cite{GengHsiao5,GengHsiaoHY},
where $a_i=c^{eff}_i+c^{eff}_{i\pm1}/N_c$
with the color number $N_c$ for $i=$odd (even) in terms of the effective Wilson coefficients $c^{eff}_i$,
defined in Refs.~\cite{Hamiltonian,ali}.
Note that $N_c$ is floating between 2 and $\infty$ in the generalized factorization
for the correction of the non-factorizable effects. In Eqs.~(\ref{amp1a})-(\ref{amp2}),
the matrix element for the annihilation of the pseudoscalar meson is defined by
\begin{eqnarray}
\label{decayconst}
\langle 0|\bar q_1 \gamma_\mu \gamma_5 q_2| P\rangle&=&if_P q_\mu\,,
\end{eqnarray}
with $f_P$ the decay constant, from which
we can obtain $\langle 0|\bar q_1 \gamma_5 q_2| P\rangle$ by using
the the equation of motion:
$-i\partial^\mu(\bar q_1\gamma_\mu q_2)=(m_{q_1}-m_{q_2})\bar q_1 q_2$ and
$-i\partial^\mu(\bar q_1\gamma_\mu\gamma_5 q_2)=(m_{q_1}+m_{q_2})\bar q_1\gamma_5 q_2$.
For the dibaryon production, the matrix elements read
\begin{eqnarray}\label{timelikeF}
\langle {\bf B}{\bf\bar B'}|\bar q_1\gamma_\mu q_2|0\rangle
&=&\bar u\bigg\{F_1\gamma_\mu+\frac{F_2}{m_{\bf B}+m_{\bf \bar B'}}i\sigma_{\mu\nu}q_\mu\bigg\}v\;,\nonumber\\
\langle {\bf B}{\bf\bar B'}|\bar q_1\gamma_\mu \gamma_5 q_2|0\rangle
&=&\bar u\bigg\{g_A\gamma_\mu+\frac{h_A}{m_{\bf B}+m_{\bf \bar B'}}q_\mu\bigg\}\gamma_5 v\,,\nonumber\\
\langle {\bf B}{\bf\bar B'}|\bar q_1 q_2|0\rangle &=&f_S\bar u v\;,\nonumber\\
\langle {\bf B}{\bf\bar B'}|\bar q_1\gamma_5 q_2|0\rangle &=&g_P\bar u\gamma_5 v\,,
\end{eqnarray}
with $u$($v$) is the (anti-)baryon spinor, where $F_{1,2}$,
$g_A$, $h_A$, $f_S$, and $g_P$ are the timelike baryonic form factors.
The amplitudes ${\cal A}_1$ and ${\cal A}_2$ now can be reduced as
\begin{eqnarray}\label{A1A2}
&&{\cal A}_1\propto 
\frac{1}{(m_{\bf B}+m_{\bf\bar B'})}\;\bar u\bigg[(m_{\bf B}+m_{\bf\bar B'})^2 g_A
+m_{B(D_s)}^2 h_A\bigg]\gamma_5 v
\,,\nonumber\\
&&{\cal A}_2\propto \frac{m_B^2}{m_b}\;\bar u (f_S +g_P \gamma_5)v\,.
\end{eqnarray}
Note that  $f_S$ and $g_P$ are not suppressed by any relations, such that
the factorization obviously works for the decay modes with ${\cal A}_2$.
Besides, the absence of $F_{1,2}$ in ${\cal A}_1$ corresponds to the conserved vector current (CVC).
However,  due to the equation of motion $F_1$ reappears as a part of $f_S$ in ${\cal A}_2$, given by
\begin{eqnarray}
f_S=n_q F_1\,,
\end{eqnarray}
with $n_q=(m_{\bf B}-m_{\bf B'})/(m_{q_1}-m_{q_2})$,
which is fixed to be 1.3~\cite{ChuaHouTsai,GengHsiao5},
presenting $30\%$ of the $SU(3)$ flavor symmetry breaking effect.
In pQCD counting rules, the momentum dependences of $F_1$ and $g_A$
can be written as \cite{Brodsky1,Brodsky2,Brodsky3}
\begin{eqnarray}\label{timelikeF2}
&&F_1=\frac{C_{F_1}}{t^2}\bigg[\text{ln}\bigg(\frac{t}{\Lambda_0^2}\bigg)\bigg]^{-\gamma}\;,
\qquad g_A=\frac{C_{g_A}}{t^2}\bigg[\text{ln}\bigg(\frac{t}{\Lambda_0^2}\bigg)\bigg]^{-\gamma}\;,
\end{eqnarray}
with $t\equiv (p_{\bf B}+p_{\bf B'})^2$, where $\gamma=2+4/(3\beta)=2.148$
with $\beta$ being the QCD $\beta$ function and $\Lambda_0=0.3$ GeV.
We note that, as the leading order expansion, $F_1$ and $g_A$ ($\propto 1/t^2$) account for 2
hard gluons, which connect to the valence quarks within the dibaryon.
In terms of  
PCAC, one obtains the relations of
\begin{eqnarray}\label{hAf}
h_A=-\frac{(m_{\bf B}+m_{\bf B'})^2}{t-m_{M}^2}g_A\,,\qquad
g_P=-\frac{m_{\bf B}+m_{\bf B'}}{m_{q_1}+m_{q_2}}\frac{m_M^2}{t-m_{M}^2}g_A\,,
\end{eqnarray}
where $m_{M}$ stands for the meson pole, while $g_P$ is related to $g_A$
 from the equation of motion.
When  $h_A$ in Eq.~(\ref{hAf}) is used for $B\to {\bf B\bar B'}$ with $t=m_B^2\gg m_M^2$,
${\cal B}(\bar B^0_{(s)}\to p\bar p)$ with a suppressed $A_1\simeq 0$
fails to explain the data by several orders of magnitude.
Similarly, ${\cal B}(\bar B^0\to \Lambda\bar p\pi^+(\rho^+))$ cannot be understood either
with $g_P$ in Eq.~(\ref{hAf})~\cite{Cheng:2001tr,ChuaHouTsai}.
We hence conclude  that
 $h_A$ and $g_P$ in Eq.~(\ref{hAf}) from PCAC at the GeV scale
are unsuitable. Recall that $F_1$ and $g_A$,
where $F_1=F_1(0)/(1-t/m_V^2)^2$ and $g_A=g_A(0)/(1-t/m_A^2)^2$~\cite{F1gA}
with the pole effects for low momentum transfer,
have been replaced by Eq.~(\ref{timelikeF2}) for the decays at the GeV scale.
It is reasonable to rewrite $h_A$ and $g_P$ to be
\begin{eqnarray}\label{hAn}
h_A=\frac{C_{h_A}}{t^2}\,,\qquad g_P=f_S\,,
\end{eqnarray}
where $h_A$ is inspired by the  relation in Eq.~(\ref{hAf}).
For  $h_A$ in Eq.~(\ref{hAf}),
since the pre-factor, $-(m_{\bf B}+m_{\bf B'})^2/t$, arises from the equation of motion,
it indicates that both $h_A$ and $g_A$ behave as $1/t^2$. Besides, at the threshold area of
$t \simeq (m_{\bf B}+m_{\bf B'})^2$, it turns out that $h_A\simeq -g_A$.
We regard
$h_A={C_{h_A}}/{t^2}$ as the modification of Eq.~(\ref{hAf}).
Consequently, PCAC is violated, $i.e.$, the axial-vector current is no more asymptotically conserved.
As a result of the $SU(3)$ flavor and $SU(2)$ helicity symmetries,
$g_P=f_S$ was first derived in Ref.~\cite{ChuaHou}, which  successfully
explained ${\cal B}(\bar B^0\to \Lambda\bar p\pi^+(\rho^+))$~\cite{ChuaHou,Geng:2011pw}.
%

In Refs.~\cite{Brodsky1,Brodsky2,Brodsky3,Geng:2011pw},
$C_{F_1}$ and $C_{g_A}$ have been derived carefully to be combined
as another set of parameters $C_{||}$ and $C_{\overline{||}}$,
which are from the chiral currents.
Here, we take the $p\bar n$ production for our description.
First, due to the crossing symmetry,
$\langle p\bar n|(\bar u d)_{V(A)}|0\rangle$ for the timelike $p\bar n$ production
and $\langle p|(\bar u d)_{V(A)}|n\rangle$ for the spacelike $n$ to $p$ transiton
are in fact identical.
Therefore, the approach of the pQCD counting rules
for the spacelike ${\bf B'\to B}$ transition is useful~\cite{Brodsky3} .
We hence combine the vector and axial-vector quark currents,
$V_\mu=\bar u\gamma_\mu d$ and $A_\mu=\bar u\gamma_\mu\gamma_5 d$,
to be the the right-handed chiral current $J^\mu_R=(V^\mu+A^\mu)/2$,
which corresponds to another set of matrix elements for the $n$ to $p$ transition:
\begin{eqnarray}\label{Gff1}
\langle p_{R+L}|J^\mu_R|n_{R+L}\rangle=
\bar u\bigg[\gamma_\mu \frac{1+\gamma_5}{2}G^\uparrow(t)+\gamma_\mu \frac{1-\gamma_5}{2}G^\downarrow(t)\bigg]u\;,
\end{eqnarray}
where the two chiral baryon states $|{\bf B}_{R+L}\rangle$
become the two helicity states
$|{\bf B}_{\uparrow+\downarrow}\rangle\equiv |{\bf B}_{\uparrow}\rangle+|{\bf B}_{\downarrow}\rangle$
in the large $t$ limit.
The new set of form factors $G^\uparrow(t)$ and $G^\downarrow(t)$ are defined as
\begin{eqnarray}\label{Gff2}
G^\uparrow(t)=e^\uparrow_{||}G_{||}(t)+e^\uparrow_{\overline{||}}G_{\overline{||}}(t)\;,\;\;
G^\downarrow(t)=e^\downarrow_{||}G_{||}(t)+e^\downarrow_{\overline{||}}G_{\overline{||}}(t)\;,
\end{eqnarray}
where
\begin{eqnarray}\label{G||}
G_{||(\overline{||})}(t) &=&
\frac{C_{||(\overline{||})}}{t^2}\bigg[\text{ln}\bigg(\frac{t}{\Lambda_0^2}\bigg)\bigg]^{-\gamma}\,,
\\
\label{Gff3}
e^{\uparrow}_{||(\overline{||})}  & = & \langle {p_{\uparrow}}|{\bf Q_{||(\overline{||})}}|{n_{\uparrow}}\rangle\;,\;\;
e^{\downarrow}_{||(\overline{||})}           =\langle {p_{\downarrow}}|{\bf Q_{||(\overline{||})}}|{n_{\downarrow}}\rangle\;,
\end{eqnarray}
which characterize the conservation of $SU(3)$ flavor and $SU(2)$ spin symmetries in the $n\to p$ transition.
Note that ${\bf Q_{||(\overline{||})}}=\sum_i Q_{||(\overline{||})}(i)$
with $i=1,2,3$ as the the chiral charge operators are coming from
$Q_{R}\equiv J^0_{R}=u_R^\dagger d_R$,
which convert one of the valence $d$ quarks in $|n_{\uparrow,\downarrow}\rangle$
to be the $u$ quark, while the converted $d$ quark can be parallel or antiparallel to the $n$'s helicity,
denoted as the subscript ($||$ or $\overline{||}$).
By comparing Eqs. (\ref{timelikeF}) and (\ref{timelikeF2}) with
Eqs. (\ref{Gff1}), (\ref{Gff2}), (\ref{G||}), and (\ref{Gff3}),
we obtain
\begin{eqnarray}
\label{form9}
C_{F_1}&=&(e^\uparrow_{||}+e^\downarrow_{||})C_{||}+(e^\uparrow_{\overline{||}}+e^\downarrow_{\overline{||}})C_{\overline{||}}\;,\nonumber\\
C_{g_A}&=&(e^\uparrow_{||}-e^\downarrow_{||})C_{||}+(e^\uparrow_{\overline{||}}-e^\downarrow_{\overline{||}})C_{\overline{||}}\;,
\end{eqnarray}
with $(e^{\uparrow}_{||},e^{\uparrow}_{\overline{||}},e^{\downarrow}_{||},e^{\downarrow}_{\overline{||}})=$
$(4/3,0,0,-1/3)$ for the $n$ to $p$ transition.
Similarly, we are able to relate $C_{F_1}$ and $C_{g_A}$ for other decay modes,
given in Table~\ref{timelikeF3}. 
However, $C_{h_A}$ in Eq.~(\ref{hAn})
only has the $SU(3)$ flavor symmetry to relate different decay modes,  given by
\begin{eqnarray}
\langle {\bf B}^i_a {\bf \bar B}'^j_{\;b}|(A_{\mu})^k_c|0\rangle=
\bar u\bigg[D d^{ijk}_{abc}+F f^{ijk}_{abc}+S s^{ijk}_{abc}\bigg]q_\mu \gamma_5 v\,,
\end{eqnarray}
where $D=C_D/t^2$, $F=C_F/t^2$, and $S=C_S/t^2$ stand for
the symmetric, anti-symmetric, and singlet form factors for $h_A$,
${\bf B}^i_a$ and ${\bf \bar B}'^j_{\;b}$ are the baryon  and anti-baryon octets,
$d^{ijk}_{abc}$, $f^{ijk}_{abc}$, and $s^{ijk}_{abc}$ are given by~\cite{TDLee}
\begin{eqnarray}
d^{ijk}_{abc}=\delta^i_b \delta^j_c \delta^k_a+\delta^i_c \delta^j_a \delta^k_b\,,\;\;
f^{ijk}_{abc}=\delta^i_b \delta^j_c \delta^k_a-\delta^i_c \delta^j_a \delta^k_b\,,\;\;
s^{ijk}_{abc}=\delta^i_b \delta^j_a \delta^k_c\,,
\end{eqnarray}
respectively.
For $\langle p\bar n|\bar u \gamma_\mu\gamma_5 d|0\rangle$,
 $(A_{\mu})^1_2=\bar u\gamma_\mu\gamma_5 d$,
we obtain $C_{h_A}=C_D+C_F$
in terms of ${\bf B}^1_3 {\bf \bar B}'^3_{\;2}=p\bar n$.
We also list $C_{h_A}$ for other decay modes in Table~\ref{timelikeF3}.
%
%
\begin{table}[t!]
\caption{The parameters $C_{F_1}$ and $C_{g_A}$ in Eq.
(\ref{timelikeF2}) are combined with $C_{||}$ and $C_{\overline{||}}$,
where the upper (lower) sign is for $C_{F_1}$ ($C_{g_A}$),
while $C_{h_A}$ consists of $C_D$, $C_F$ and $C_S$.} \label{timelikeF3}
\begin{center}
\begin{tabular}{|c|c|c|}
\hline
matrix element   &$C_{F_1}$($C_{g_A}$)&$C_{h_A}$\\\hline
$\langle p\bar p|(\bar u u)|0\rangle$&$\frac{5}{3}C_{||}\pm\frac{1}{3}C_{\overline{||}}$&
$C_D+C_F+C_S$\\
$\langle p\bar p|(\bar d d)|0\rangle$&$\frac{1}{3}C_{||}\pm\frac{2}{3}C_{\overline{||}}$&
$C_S$\\
$\langle p\bar p|(\bar s s)|0\rangle$&0&
$C_D-C_F+C_S$\\
$\langle p\bar n|(\bar u d)|0\rangle$&$\frac{4}{3}C_{||}\mp\frac{1}{3}C_{\overline{||}}$&
$C_D+C_F$\\
$\langle \Sigma^-\bar \Sigma^0|(\bar d u)|0\rangle$&$\frac{1}{3\sqrt 2}(5C_{||}\pm C_{\overline{||}})$&
$\sqrt 2 C_F$\\
$\langle  \Lambda\bar \Lambda|(\bar u u)|0\rangle$&$\frac{1}{2}C_{||}\pm\frac{1}{2}C_{\overline{||}}$&
$\frac{1}{3}C_D+C_S$\\
$\langle  \Lambda\bar \Lambda|(\bar d d)|0\rangle$&$\frac{1}{2}C_{||}\pm\frac{1}{2}C_{\overline{||}}$&
$\frac{1}{3}C_D+C_S$\\
$\langle  \Lambda\bar \Lambda|(\bar s s)|0\rangle$&$C_{||}$&
$\frac{4}{3}C_D+C_S$\\
$\langle  \Lambda\bar p|(\bar s u)|0\rangle$            &$-\sqrt\frac{3}{2}C_{||}$&
$-\frac{1}{\sqrt 6}(C_D+3C_F)$\\
$\langle  \Sigma^0\bar p|(\bar s u)|0\rangle$           &$\frac{-1}{3\sqrt 2}(C_{||}\pm 2C_{\overline{||}})$&
$\frac{1}{\sqrt 2}(C_D-C_F)$\\
\hline
\end{tabular}
\end{center}
\end{table}

\section{Numerical analysis}
For the numerical analysis, the CKM matrix elements and
the quark masses are taken from the particle data group (PDG)~\cite{pdg},
where $m_b=4.2$ GeV. The decay constants in Eq.~(\ref{decayconst}) are given by
\cite{Aubin:2005ar,Na:2012kp}
\begin{eqnarray}
(f_B,\,f_{B_s},\,f_{D_s})=(190,\,225,\,250)\; \text{MeV}\,.
\end{eqnarray}
For the parameters in Table~\ref{timelikeF3}, we refit $C_{||}$ and $C_{\overline{||}}$
by the approach of Ref.~\cite{NF_GengHsiao}
with the data of ${\cal B}(\bar B^0_{(s)}\to p\bar p)$, ${\cal B}(D^+_s\to p\bar n)$,
${\cal B}(\bar B^0\to n\bar p D^{*+})$, and ${\cal B}(\bar B^0\to \Lambda\bar p \pi^+)$,
while $C_D$, $C_F$ and $C_S$ are newly added in the fitting.
Note that the OZI suppression makes $\langle p\bar p|(\bar s s)|0\rangle=0$, which results in
$C_S=C_F-C_D$. With $N_c=2$ fixed in $a_i$ as the best fit, 
the parameters are fitted to be
\begin{eqnarray}\label{para}
(C_{||},\,C_{\overline{||}})&=&(-102.4\pm 7.3,\,210.9\pm 85.2)\,\text{GeV}^{4}\,,\nonumber\\
(C_D,\,C_F)&=&(-1.7\pm 1.6,\,4.2\pm 0.7)\,\text{GeV}^{4}\,.
\end{eqnarray}
As shown in Table~\ref{branchingratios},
we can reproduce the data of $\bar B_{(s)}^0\to p\bar p$ and $D_s^+\to p\bar n$.
In addition,  we predict the branching ratios of
$\bar B^0_{(s)}\to \Lambda\bar \Lambda$,
$B^-\to (\Lambda\bar p, \Sigma^0\bar p)$, and
$B^-\to (n\bar p, \Sigma^- \bar\Sigma^0)$ in Table~\ref{branchingratios}.
\begin{table}[b!]
\caption{ The branching ratios of $B_{(s)}\to{\bf B\bar B'}$ ($D_s\to{\bf B\bar B'}$) decays in units of $10^{-8}$ ($10^{-3}$), where
the uncertainties arise from
the time-like baryonic $0\to {\bf B \bar B'}$ form factors.}\label{branchingratios}
\begin{center}
\begin{tabular}{|c|c|c|}
\hline
decay mode               &our result&data\\\hline
$\bar B^0\to p\bar p$ & $1.4^{+0.5}_{-0.5}$&$1.47^{+0.71}_{-0.53}$~\cite{Aaij:2013fta}\\
$\bar B_s^0\to p\bar p$& $3.0^{+1.5}_{-1.2}$& $2.84^{+2.20}_{-1.69}$~\cite{Aaij:2013fta}\\
$D_s^+\to p\bar n$ & $1.3^{+13.2}_{-\;\;1.3}$&$1.30^{+0.38}_{-0.39}$~\cite{Athar:2008ug}\\
$B^-\to n\bar p$ & $3.2^{+6.9}_{-3.0}$&---\\
$B^-\to \Lambda\bar p$ & $3.5^{+0.7}_{-0.5}$&$<32$~\cite{Tsai:2007pp}\\
$\bar B^0\to \Lambda\bar \Lambda$ & $0.3^{+0.2}_{-0.2}$&$<32$~\cite{Tsai:2007pp}\\
$\bar B^0_s\to \Lambda\bar \Lambda$ & $5.3^{+1.4}_{-1.2}$&---\\
$B^-\to \Sigma^0 \bar p$ & $5.3^{+3.8}_{-2.7}$&---\\
$B^-\to \Sigma^- \bar\Sigma^0$ & $9.6^{+4.0}_{-3.3}$&---\\
\hline
\end{tabular}
\end{center}
\end{table}

\section{Discussions and Conclusions}
When the axial-vector current is not asymptotically conserved, we can evaluate
the two-body baryonic $B_{(s)}$ and $D_s$ decays with the annihilation mechanism
 to explain the data. In particular, the experimental values of ${\cal B}(\bar B^0_{(s)}\to p\bar p)$
and ${\cal B}(D_s^+\to p\bar n)$ can be reproduced.
It is the violation of  PCAC that makes ${\cal B}(D_s^+\to p\bar n)$ to be of order $10^{-3}$,
which was considered as the consequence of the long-distance contribution in Ref.~\cite{Chen:2008pf}.
With $m_{D_s}\simeq m_p+m_{\bar n}$, the amplitude of ${\cal A}_1(D_s^+\to p\bar n)$
from Eq. (\ref{A1A2}) is in fact proportional to $\bar u(g_A+h_A)v$.
Instead of  $h_A=-g_A$ from PCAC in Eq.~(\ref{hAf}) with t=$m_{D_s}^2$,
our approach with $h_A=-0.7 g_A$ shows that the $30\%$ broken effect of PCAC suffices
to reveal  ${\cal B}(D_s^+\to p\bar n)$.
As seen from Table~\ref{timelikeF3},
$C_{h_A}=C_D+C_F$ for the $p\bar n$ production with the uncertainties
fitted in Eq.~(\ref{para}) has the solutions of $h_A=0$ to $h_A=-g_A$,
which allows ${\cal B}(D_s^+\to p\bar n)=(0-16)\times 10^{-3}$.
With the OZI suppression of $\langle p\bar p|(\bar s s)|0\rangle=0$,
which eliminates ${\cal A}_2$, the decay of $\bar B^0_{s}\to p\bar p$
is the same as that of   $\bar B^0\to p\bar p$ to be the first type.
In contrast with $D_s^+\to p\bar n$, since  ${\cal A}_1(\bar B^0_{(s)}\to p\bar p)$
$\propto m_B^2[(\frac{m_p+m_{\bar p}}{m_{B}})^2 g_A+h_A]\bar u\gamma_5 v$
with a suppressed $g_A$ contribution at the $m_B$ scale, the decay branching ratios
are enhanced by $h_A$ with $m_B^2$.
Similarly, being of the first type, our predicted results for
${\cal B}(\bar B^0\to \Lambda\bar \Lambda)$,
${\cal B}(B^-\to n\bar p)$ and
${\cal B}(B^-\to \Sigma^- \bar\Sigma^0)$
can be used to  test the violation of PCAC at the GeV scale.

On the contrary,
${\cal B}(B^-\to \Lambda(\Sigma^0)\bar p)$
and ${\cal B}(\bar B^0_s\to \Lambda\bar \Lambda)$ are primarily contributed from ${\cal A}_2$.
Similar to  the theoretical relation between  $B^-\to p\bar p \ell \bar \nu$~\cite{semi}
and  $B\to p\bar p M$, which are associated with
the same form factors in the $B$ to $\bf B\bar B'$ transition,
resulting in the first observation of the semileptonic baryonic $B$ decays~\cite{Tien:2013nga},
there are connections between the two-body $B^-\to \Lambda(\Sigma^0)\bar p$ and $\bar B^0_s\to \Lambda\bar \Lambda$
and three-body $\bar B^0\to \Lambda\bar p\pi^+$ and $B\to \Lambda\bar \Lambda K$ decays
with the same form factors via the (pseudo)scalar currents.
As a result, without  PCAC, the observations of these two-body modes can serve as the test of the factorization,
which accounts for the short-distance contribution.
Note that the recent work by fitting $\bar B^0\to p\bar p$
with the non-factorizable contributions leads
${\cal B}(\bar B^0_s\to p\bar p)$ and ${\cal B}(\bar B^0\to \Lambda\bar \Lambda)$
to be nearly zero~\cite{diagramic2}, which are clearly different from our results.

In sum, we have proposed that, based on the factorization, the annihilation mechanism
can be applied to all of the two-body baryonic $B_{(s)}$ and $D_s$ decays,
which indicates that the hypothesis of PCAC is violated at the GeV scale.
With the modified timelike baryonic form factors via the axial-vector currents,
we are able to explain ${\cal B}(\bar B^0_{(s)}\to p\bar p)$
and ${\cal B}(D_s^+\to p\bar n)$
of order $10^{-8}$ and  $10^{-3}$, respectively.
For the decay modes that have the contributions from the (pseudo)scalar currents,
they have been predicted as
${\cal B}(B^-\to \Lambda\bar p)=(3.5^{+0.7}_{-0.5})\times 10^{-8}$,
${\cal B}(B^-\to \Sigma^0 \bar p)=(5.3^{+3.8}_{-2.7})\times 10^{-8}$, and
${\cal B}(\bar B^0_s\to \Lambda\bar \Lambda)=(5.3^{+1.4}_{-1.2})\times 10^{-8}$,
which can be used to test the annihilation mechanism.
Besides, the branching ratios of
$\bar B^0\to \Lambda\bar \Lambda$,
$B^-\to n\bar p$, and
$B^-\to \Sigma^- \bar\Sigma^0$, predicted to be
$(0.3,\,3.2,\,9.6)\times 10^{-8}$,
can be viewed  as the test of PCAC, which are accessible to the experiments at LHCb.

\section*{ACKNOWLEDGMENTS}
We thank Professor H.Y. Cheng and Professor C.K. Chua for discussions.
This work was partially supported by National Center for Theoretical
Sciences,  National Science Council
 NSC-101-2112-M-007-006-MY3) and National Tsing Hua
University~(103N2724E1).

\end{document}